\documentclass[a4paper]{article}

\usepackage{INTERSPEECH2020}
\usepackage{amsmath,graphicx}
\usepackage{float}
\usepackage{graphicx}
\usepackage{tabularx}
\usepackage{tabulary}
\usepackage{xcolor}
\usepackage{paralist}
\usepackage{units}
\usepackage{microtype}
\usepackage{multirow}
\usepackage[multiple]{footmisc}
\usepackage{enumitem}
\usepackage{subcaption}
\usepackage{flushend}

\title{Prosody Transfer in Neural Text to Speech \\ Using Global Pitch and Loudness Features}

\name{Siddharth~Gururani$^1$\sthanks{\hspace{0.15cm}The work was done while S. Gururani was interning at Electronic Arts.} , Kilol~Gupta$^2$, Dhaval~Shah$^2$, Zahra~Shakeri$^2$, Jervis~Pinto$^2$}

\address{$^1$ Music Informatics Group, Georgia Institute of Technology, Atlanta, GA, USA \\
    $^2$ EADP Data \& AI, Electronic Arts, Redwood City, CA, USA}
\email{siddgururani@gatech.edu, \{kgupta,dshah,zshakeri,jpinto\}@ea.com}

\begin{document}

\maketitle
\begin{abstract}
This paper presents a simple yet effective method to achieve prosody transfer from a reference speech signal to synthesized speech. The main idea is to incorporate well-known acoustic correlates of prosody such as pitch and loudness contours of the reference speech into a modern neural text-to-speech (TTS) synthesizer such as Tacotron2 (TC2). 
More specifically, a small set of acoustic features are extracted from reference audio and then used to condition a TC2 synthesizer.
The trained model is evaluated using subjective listening tests and a novel objective evaluation of prosody transfer is proposed. Listening tests show that the synthesized speech is rated as highly natural and that prosody is successfully transferred from the reference speech signal to the synthesized signal.
\end{abstract}
\section{Introduction}
\label{sec:intro}

In this paper, we address the task of developing a text-to-speech (TTS) system with the following two properties: a)~The range of prosody in the synthesized dialogue must encompass a large range of human conversation, from neutral expression to extremely emotional, \emph{while always sounding perfectly natural}, and b) The user must be able to easily control prosody in synthesized speech. Here, prosody refers to the variation of several speech related phenomena such as intonation, stress, rhythm and style of the speech~\cite{skerry2018towards}. Such expressive TTS (ETTS) systems are desirable for content creators and application developers for whom the listener's enjoyment and immersiveness of the speech content is of primary importance. Although the desire for ETTS is not new~\cite{theune2006generating}, initial TTS systems lacked the ability to produce sufficiently natural neutral speech, which is a pre-requisite since human conversation is typically more neutral than expressive.

Recent progress in deep learning architectures has significantly improved the naturalness of TTS systems \cite{gibiansky2017deep, shen2018natural}. However, the style of the speech strongly depends on the training dataset, with the model often learning a typically neutral prosody \cite{wu2018rapid}. Due to this problem there has been much recent interest in developing models which can encode prosody. The resulting ETTS system would still need to be easily controllable, preferably by the creator of the text itself, who knows the target prosody.

We propose a simple yet effective ETTS system in which a domain expert can easily transfer the target prosody from one audio clip to another. In this system, the user specifies text and selects a reference speech signal. The output is synthesized speech for the input text using the reference speech's prosody. 
More specifically, we extract a set of acoustic features from a reference speech signal that contains the desired prosody. Next, we use the extracted features, which encode key prosodic components (e.g., intonation, arousal), to condition a vanilla Tacotron2 (TC2) synthesizer~\cite{shen2018natural}. The resulting log-mel-spectrogram has the desired prosody while still sounding natural. An additional advantage of our system is its low resource requirements.
Our main contributions are summarized below:
\begin{compactenum}
\item We demonstrate the utility of the pitch contour and root-mean-square (RMS) energy curve for generating expressive speech. We show that the prosody of a desired reference can be transferred to a synthesized example using as few as 7 features.
\item We use this compact reference ``encoder'' as conditioning inputs for the TC2 model. We denote the resulting model as GS-TC2 (Global Summary statistics). We demonstrate the transfer of prosody from the reference signal to the final synthesized speech signal, with a small decrease in naturalness. 
\item Finally, prosody transfer and ETTS is a relatively recent endeavor. We contribute to this effort by extending the framework for evaluating prosody transfer with new objective evaluation metrics.
\end{compactenum}

\begin{figure*}
    \centering
    \includegraphics[width=\linewidth]{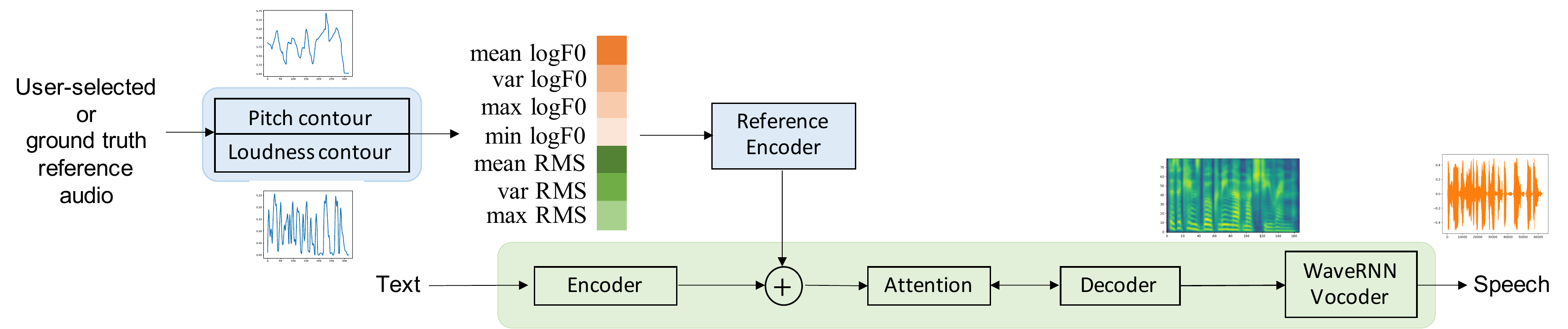}
\caption{Model architecture for prosody transfer. The reference encoder is a linear transformation of summary statistic features which is followed by element-wise sum with each output of the text encoder. During training, prosody features are obtained from ground truth audio while during inference, they are extracted from reference speech.}
    \label{fig:graphical-depiction-of-training-and-inference}
  \end{figure*}
  
\section{Related Work}
\label{sec:related}

Most research methods in ETTS modify an existing TTS framework to incorporate expressivity. These frameworks have traditionally leveraged hidden Markov models (HMM) for parametric speech synthesis~\cite{zen2007hmm}. One early approach was to use supervised learning with explicit expression labels~\cite{schroder2010synthesis}. The primary difficulty here is that annotating speech with `emotion' or `expression' is subjective and expensive. Eyben et al.~\cite{eyben2012unsupervised} proposed an approach involving unsupervised clustering of acoustic features and using cluster assignments as context for the generation process of average-expression speech synthesis.

More recently, neural TTS frameworks such as TC2~\cite{shen2018natural} are adapted for ETTS by conditioning the synthesizer with a learned embedding of prosody. Skerry-Ryan et al.~\cite{skerry2018towards} propose learning a latent space for prosody by encoding the speech signal into a lower dimensional embedding. The learned prosody embedding is used as a global conditioning vector for Tacotron's decoder~\cite{wang2017tacotron}. Wang et al.~\cite{wang2018style} devised a method to learn categories of prosody styles or `global style tokens' in an unsupervised manner. They jointly learn style embeddings along with the TC2 model by adding the style embeddings to the text encoder states. 

Similar to the global style tokens approach, Azukawa et al.~\cite{akuzawa2018expressive} conditioned the encoder of VoiceLoop~\cite{taigman2018voiceloop} with a latent variable containing global speech characteristics. The latent space is learned using a Variational Autoencoder (VAE). Hsu et al.~\cite{hsu2018hierarchical} formulated a conditional generative model for speech where a Gaussian Mixture VAE is utilized to infer a latent code (prosody or style) as well as an observed code (speaker identity). These codes are used to condition the decoder of a TC2 model. The results show significant improvement in Mean Opinion Scores (MOS) of synthesized speech over previous approaches. Moreover, Wan et al.~\cite{wan2019chive} proposed a hierarchical conditional VAE which utilizes both prosodic features such as pitch, loudness and phoneme duration, as well as linguistic features to learn a prosody embedding.In this prosody embedding model, the features are encoded at different rates using clockwork RNNs. This differs from other VAE methods in that the prosody embedding is learned from acoustic and linguistic features as opposed to the spectrogram.


While the above methods proposed ways to control or transfer prosody from a reference speech, a few researchers are working on ways to enable fine-grained control of prosodic features. Lee and Kim~\cite{lee2019robust} proposed a method to learn variable-length prosody embeddings and conditioning a TC2 model. The approach uses attention to align and downsample the prosody embeddings learned from the audio to match the number of time-steps in the encoder or the decoder. Klimkov et al.~\cite{Klimkov2019} utilized a similar strategy for fine-grained prosody transfer using phoneme aligned prosody features such as pitch, intensity, and duration which are encoded by a reference encoder. The prosody embeddings are utilized by a TC2-based system to generate mel-spectrograms of expressive speech. In a more recent work, Valle et al.~\cite{valle2019mellotron} propose ``Mellotron" that explicitly conditions the TC2 model on continuous rhythm and pitch contours to achieve fine-grained control of rhythm and pitch of synthesized speech.

\section{Proposed Method}
\label{sec:method}

Our approach leverages prior research on human speech and signal processing which showed that pitch and loudness could encode significant amounts of speech prosody \cite{theune2006generating}. Our hypothesis is that features derived from pitch and loudness contours of a speech signal can be combined with neural TTS to enable prosody transfer. We test this hypothesis by incorporating these prosodic features into a standard TC2 TTS model using global conditioning, previously shown to be effective \cite{hsu2018hierarchical, wang2018style}. A key aspect of our method is that we do not learn prosody or style features from high-dimensional spectrogram representations of audio \cite{skerry2018towards, hsu2018hierarchical}. Rather, we convert the reference speech signal into 1-D time-series features such as pitch contours and frame-level RMS energy.

The main advantage of our approach is that the low-dimensional features allow us to use a compact reference encoder, with a very small number of additional parameters. Specifically, we only need to add a single linear projection layer that maps low dimensional prosody features to 512 dimensions to match the dimensionality of the TC2 conditioning site. Another potential advantage of the low-dimensional prosody encoding is that it can help the model disentangle text and speaker attributes from prosodic ones. These attributes distinguish our method from state-of-the-art prosody transfer models \cite{hsu2018hierarchical}, which involve both complex reference speech encoders requiring longer training as well as larger datasets due to the model complexity, while still experiencing difficulties from entangled representations since they rely on a high-dimensional (i.e., spectrogram) representation of the reference. Furthermore, approaches involving VAEs can be difficult to train whereas our model can be trained on a single V100 NVIDIA GPU using a variant of a public TC2 implementation.\footnote{https://github.com/NVIDIA/tacotron2}

To the best of our knowledge, there has been very little research investigating the effectiveness of using pitch and loudness features to condition a neural synthesizer. The closest approach to our work is the method proposed by Wan et al.~\cite{wan2019chive} which uses similar acoustic features in addition to linguistic features. However, the model is quite large and has significant implementation complexity involving clockwork RNNs, seq2seq and variational inference. In comparison, we propose a straightforward modification to a vanilla TC2 model with low-dimensional, well-studied acoustic attributes of human speech, namely fundamental frequency (F0) and energy (RMS).
Note that the approach taken in Mellotron \cite{valle2019mellotron} is intuitively similar to our approach, but has major implementation differences. We utilize loudness characteristics to encode performances with varying loudness, from "quiet" speech (e.g., whispers) to "loud" (e.g., shouting), which Mellotron does not. Additionally, Mellotron enables fine-grained control whereas our approach transfers global prosody characteristics to the output speech.
Our method of generating expressive speech via prosody transfer is graphically depicted in Figure~\ref{fig:graphical-depiction-of-training-and-inference}.

\subsection{Extracting Prosody Features}


Prosody refers to the variation of pitch and stress in speech, among other factors. In order to inform the TTS model about the user's desired prosody for synthesized speech, we let the user specify a reference speech signal containing the desired prosody. From this reference signal, we extract two 1-D signals (time series), one for the fundamental frequency (F0) and the second for loudness. We represent loudness as the RMS of overlapping frames of audio. This is a fast and straightforward computation. For fundamental frequency (i.e., pitch contour), we use a python implementation of the Kaldi speech recognition toolkit~\cite{Povey_ASRU2011}. Kaldi uses a normalized cross-correlation function to compute the  pitch contour. We refer readers to~\cite{ghahremani2014pitch} for details about the pitch tracking algorithm. The time-series is a smoothed logF0 value for each audio frame. Unvoiced frames are set to 0 in the pitch contour based on a threshold on the RMS that is computed offline.

Given logF0 and RMS curves, our approach for computing the conditioning features consists of extracting ``global statistics'' (mean, variance, maximum, minimum) over the two time series. This method, denoted as GS, gives us a total of 7 features.\footnote{We exclude the minimum feature for RMS since it is typically zero and not informative.} In the next section, we demonstrate how these simple features lead to the effective transfer of prosody from a reference speech signal to synthesized speech signal with small decrease in speech naturalness.

\section{Experiments and Results}
\label{sec:exp}

This section contains the key empirical findings of the paper on a single speaker dataset of expressive speech.

\subsection{Dataset and Preprocessing}

Our proprietary dataset consists of approximately 10 hours of professionally recorded, female, single-speaker expressive speech in English, spoken with a British accent. There are approximately 10000 (text, utterance) pairs in the dataset. The recorded lines cover a large variety of expressions including neutral, joyful, angry, sarcastic and urgent.\footnote{These descriptors were gathered from listeners during informal focus groups for the sole purpose of subjective analysis of the expressivity in the dataset.} Our training and inference do not leverage any form of expression annotations.

We set aside 100 utterances for validation and 100 utterances for testing. All our evaluations are done on synthesized text from the test set. 
We downsample the original audio recorded at \unit[48]{kHz} to \unit[16]{kHz}. The mel-spectrograms are computed using 80 mel bins, window size of \unit[50]{ms} and hop size of \unit[12.5]{ms}. Mel-spectrogram magnitudes are converted to the log-scale and scaled to a range of $[-4,4]$. We use the same window-size and hop-size for loudness and pitch contour computation. We apply an RMS threshold of $5e^{-3}$ as a proxy for voicing detection. For the text input, we use a character set size of 37 comprising of lower-case alphabets and normalized punctuation.
We also perform standard normalization on the 7 global statistical features.


As previously described, we experiment with conditioning TC2 with prosody features. 
The GS-TC2 model uses the same procedure for training as the standard TC2. We use a learning rate of $1e^{-3}$ and batch size 32 and run the training for approximately $200,000$ steps after which we observe that the models have converged.
During training, the conditioning inputs are obtained from the ground-truth audio recording for the input text. During inference, a reference can be chosen by an end-user based on the desired prosody. This reference can come from the training, validation or test set. Finally, we use a variant of the WaveRNN vocoder to generate waveforms from predicted log-mel-spectrograms~\cite{kalchbrenner2018efficient}. The vocoder is trained separately on ground-truth log-mel-spectrograms from training data.

\subsection{Results and Discussion}
We evaluate our models for prosody transfer using both subjective and objective metrics. Additionally, we perform MOS evaluations for speech naturalness. 

\subsubsection{Subjective Evaluations}
We begin with a MOS evaluation of speech naturalness. We use text from the test set and generate speech using the baseline TC2 and the GS-TC2 model.
We also evaluate the naturalness of ground truth (GT) recorded test audio. Note that for the GS-TC2
model, we use the corresponding audio's ground truth prosody features for conditioning. 
For each MOS test, a total of 20 listeners consisting of English speakers were selected. Each listener was presented with 9 audio files and was asked to rate the naturalness of these files on a scale of 1-5 with 0.5 increments.
The results from the MOS evaluations are shown in Table~\ref{tab:MOS}.
The addition of the prosody transfer module results in a small decrease in the naturalness of the TC2 model. We attribute this to the fact that using reference speech samples with extreme prosody (loudness and pitch) result in unwanted artifacts in the model output~\cite{theune2006generating}. This issue is not frequently observed in the TC2 model’s outputs because the TC2 model converges to the dominant prosody in the speech dataset, which is a neutral tone.\footnote{Our listening test was also accompanied by a survey. Most listeners pointed out that the presence of uncommon proper nouns or entity names caused them to give lower ratings even for ground truth recordings. We hypothesize the large standard deviations are a result of this issue.}


\begin{table}[]
  \centering
  \begin{tabular}{|l|c|c|c|}
  	\hline
 	 Model & GT & TC2  & GS-TC2 \\ \hline
 	 MOS Score  & $4.13 \pm 1.01$ & $ 3.76 \pm 1.06 $ 
  	& $3.62 \pm 1.07  $ \\ \hline
  	\multicolumn{3}{|l|}{Prosody Transfer Score}  & $1.04 \pm 1.81$ \\ 	
  	\hline
  \end{tabular}
  \caption{MOS Naturalness scores for the models and preference score for prosody transfer. Higher is better.}
  \label{tab:MOS}
  \end{table}

Next, we evaluate prosody transfer from the reference speech to the synthesized speech for an arbitrary line of text. This is a side-by-side test comparing synthesized speech from two models which was first proposed by Skerry-Ryan et al.~\cite{skerry2018towards}. We ask listeners to rate which of the two synthesized samples is more similar to a reference speech sample. Specifically, we construct triplets $(A, X, Y)$ from the test set. $A$ is the reference speech sample, $X$ and $Y$ are utterances synthesized using the same text but using TC2 and GS-TC2 respectively.
For this test, a total of 27 listeners consisting of English speakers were selected.
Each listener was presented with 6 triplets and asked to rate which synthesized waveform matches more closely to the reference in terms of prosody, on a Likert scale of -3 to 3 with increments of 1. Here,
-3 implies $X$ is closer to $A$, +3 implies $Y$ is closer to $A$, and 0 implies that there is no difference.
Table~\ref{tab:MOS} shows the results of this listening test. It can be inferred that the GS-TC2 model successfully transfers prosody to synthesized audio compared to baseline TC2.

  
\begin{figure*}
\begin{subfigure}{.25\textwidth}
  	\centering
  \includegraphics[width=1.1\linewidth]{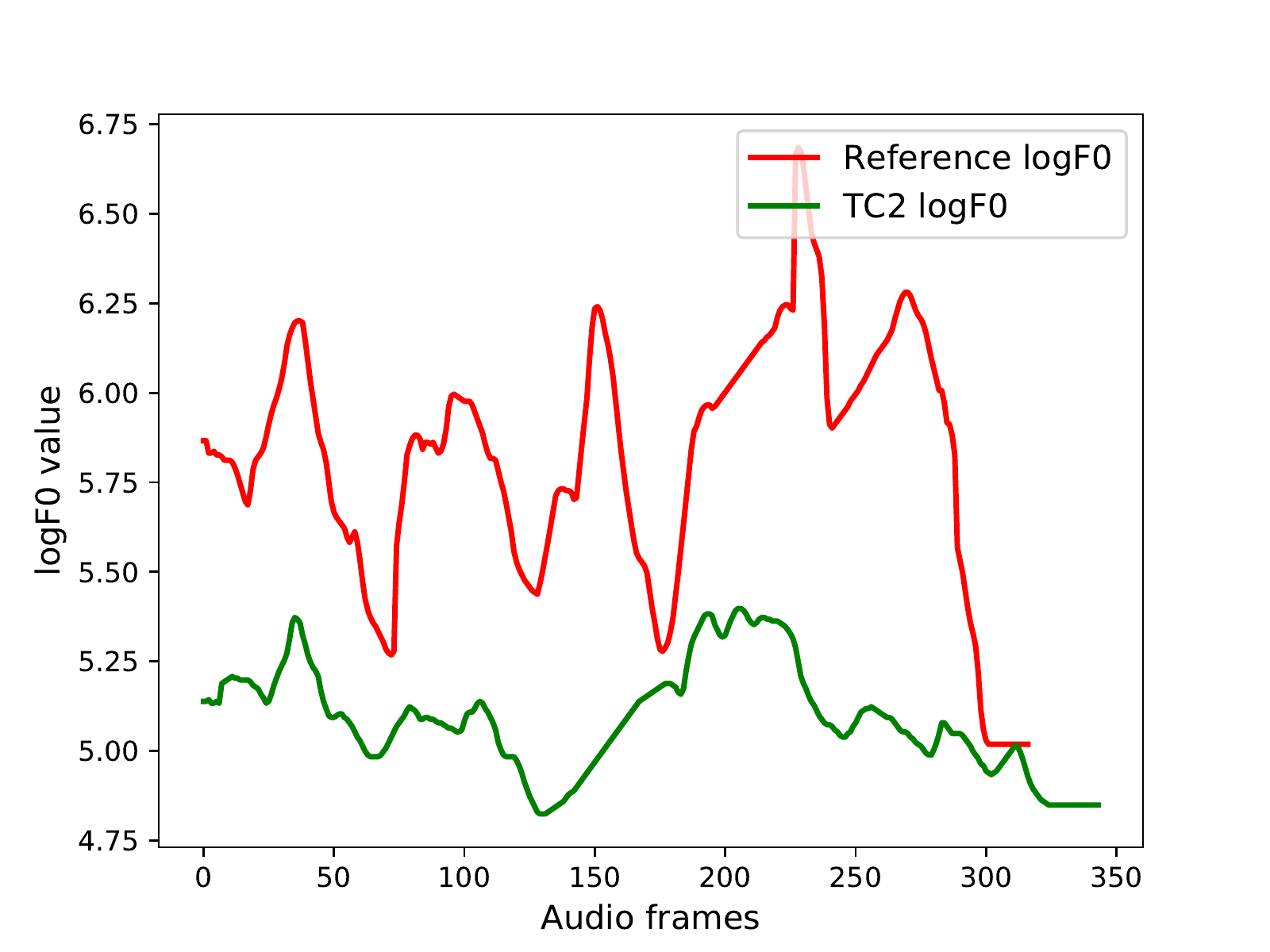}
  \caption{}
  \label{fig:sfig1}
\end{subfigure}%
\begin{subfigure}{.25\textwidth}
  \centering
  \includegraphics[width=1.1\linewidth]{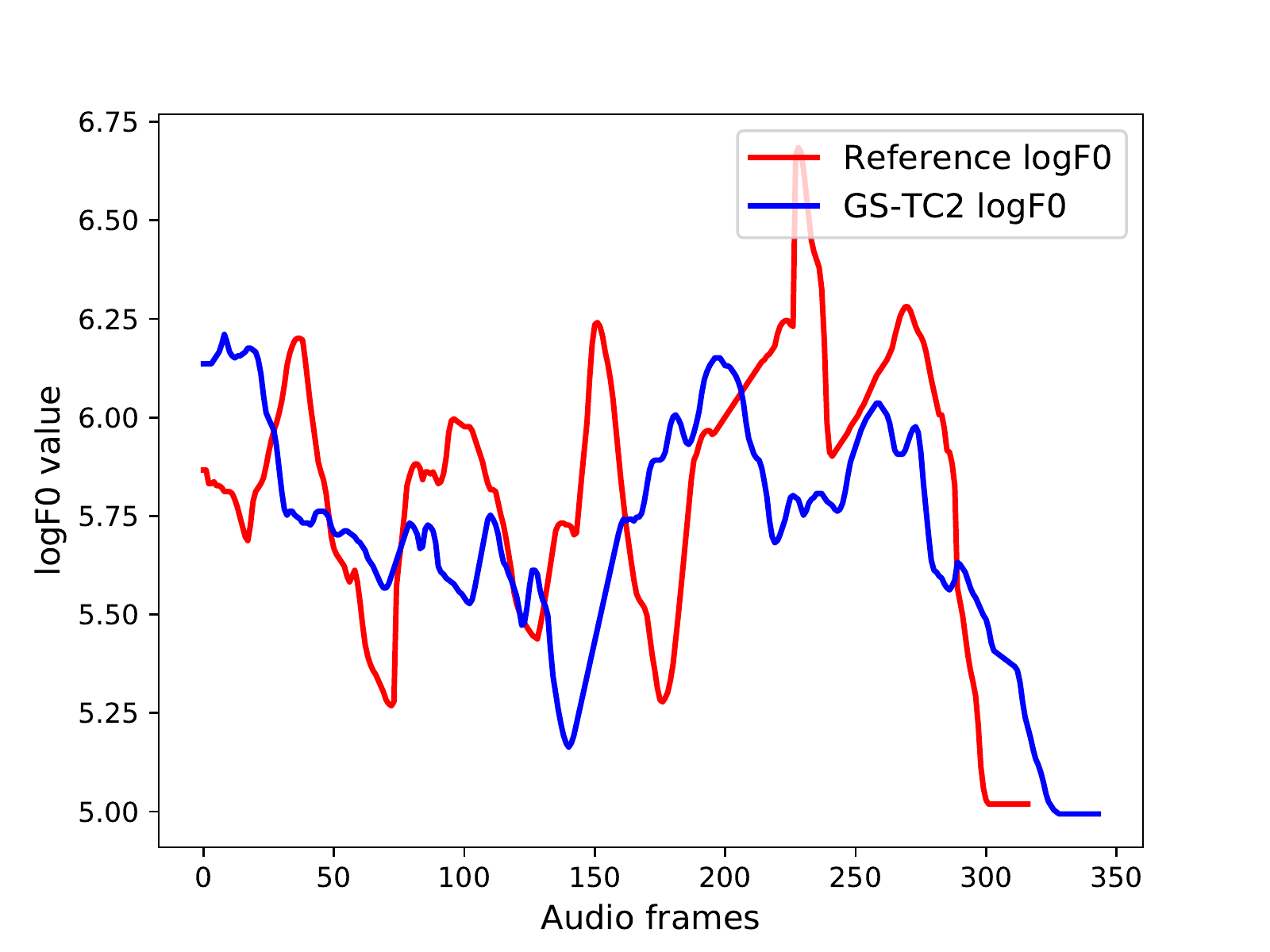}
  \caption{}
  \label{fig:sfig2}
\end{subfigure}
    \begin{subfigure}{.25\textwidth}
  	\centering
  \includegraphics[width=1.1\linewidth]{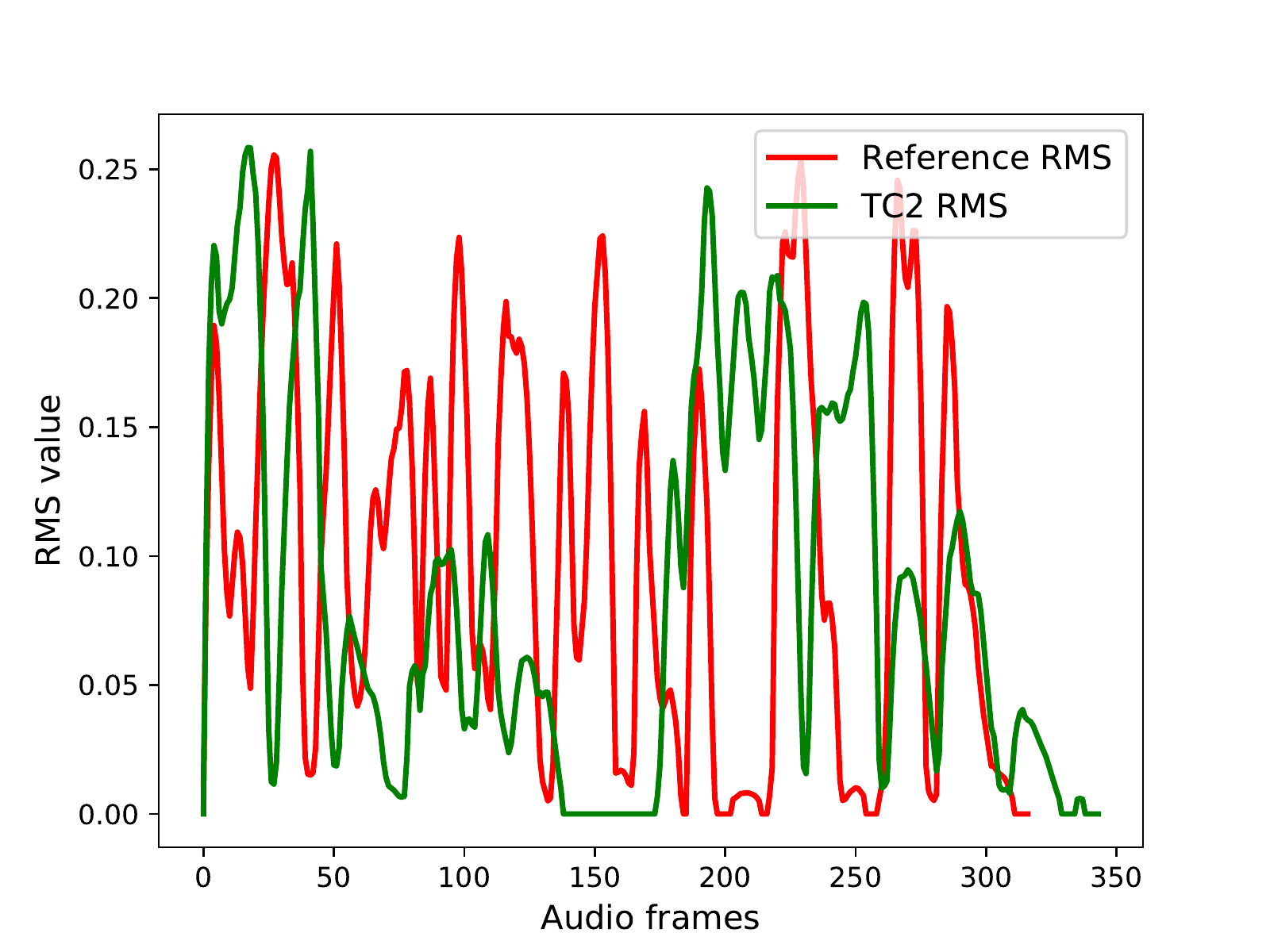}
  \caption{}
  \label{fig:sfig1}
\end{subfigure}%
\begin{subfigure}{.25\textwidth}
  \centering
  \includegraphics[width=1.1\linewidth]{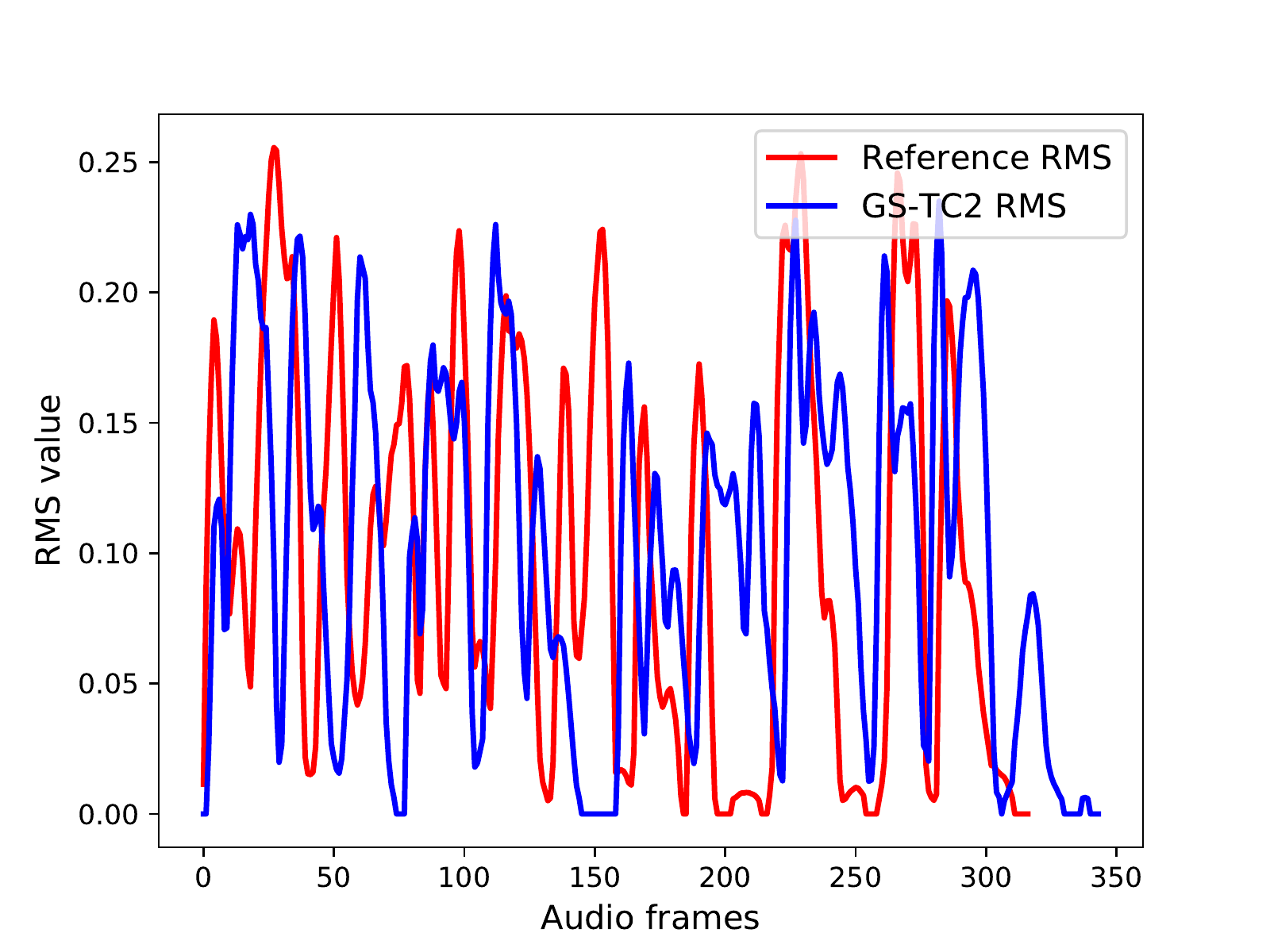}
  \caption{}
  \label{fig:sfig2}
\end{subfigure}
\caption{Example depicting logF0 and RMS contours for two synthesized waveforms (from the same transcript) using TC2 (green) and GS-TC2 (blue) compared to GT reference waveform (red - different transcript). It can be seen that the logF0 and RMS statistics of the waveform generated using GS-TC2 are closer to the statistics of the reference waveform.}
\label{fig:example_curves}
\end{figure*}
\vspace{-0.3\baselineskip}
\subsubsection{Objective Evaluations}
Our goal is to identify objective metrics that correlate strongly with subjective metrics. Since a standard set of objective metrics for prosody transfer does not exist, another contribution of this work is introducing two metrics for this purpose.
Our objective evaluations consist of measuring the prosody transfer performance of GS-TC2 compared to TC2 via:
\begin{compactenum}
	\item  Computing the cosine distance between GS Pitch and RMS features of the ground-truth recorded reference audio and the synthesized speech.

	\item Computing the dynamic time warping (DTW) distance~\cite{muller2007dynamic} between the pitch and RMS series of the ground-truth recorded reference audio and the synthesized speech.
\end{compactenum}

\begin{figure} [th!]
  \centering
  \includegraphics[width=0.9\linewidth]{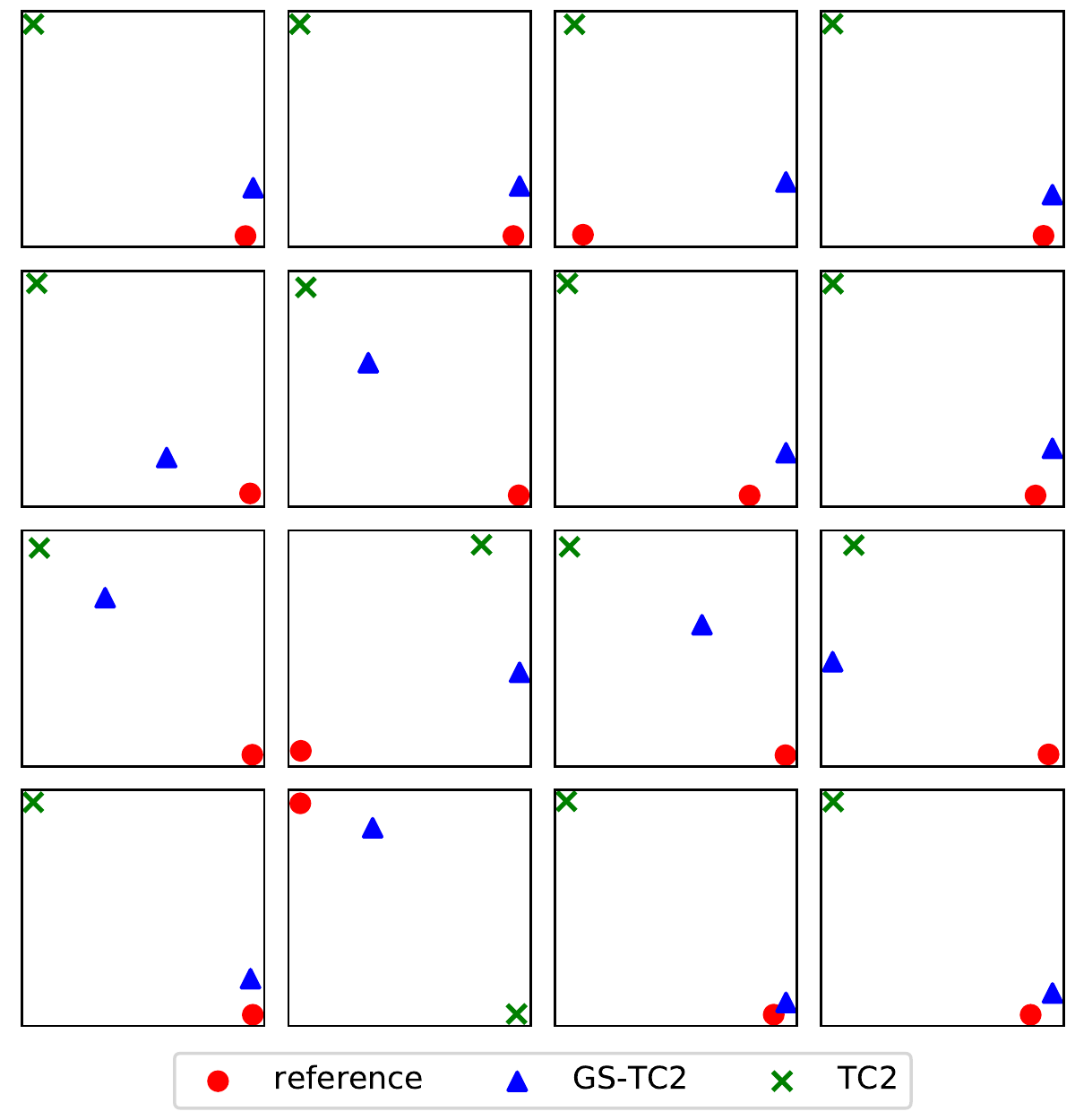}
\caption{Scatter plots showing the relative distance of pitch and loudness features of randomly chosen samples synthesized by GS-TC2 and TC2 from a reference utterance. The 7-dimensional features are projected onto 2 dimensions using t-SNE. The GS-TC2 model clearly wins 14 out of 16 times in this trial.}
  \label{fig:scatter-plot}
\end{figure}

In this experiment, for each text in the test set (consisting of 100 examples), a reference is selected uniformly at random from a set of 100 references that contain non-neutral prosodies and the text is synthesized using the GS-TC2 model, conditioned on that reference. We conducted 50 Monte Carlo runs in all cases. 
Table~\ref{tab:obj} shows the results for objective evaluation. It can be seen that GS-TC2 can synthesize speech with GS features closer to the ground-truth reference audio compared to baseline TC2. It can also be observed that GS-TC2 is more successful in transferring the reference pitch contour information to the synthesized speech compared to TC2, whereas, in terms of RMS, the DTW distance is similar for GS-TC2 and TC2. 

Figures~\ref{fig:example_curves} and~\ref{fig:scatter-plot} visualize the strong performance of the GS-TC2 model compared to TC2. Figure~\ref{fig:example_curves} shows an example of logF0 and RMS contours for two synthesized utterances using TC2 and GS-TC2 compared to the ones for the reference utterance. It can be seen that the contours for the synthesized waveform using GS-TC2 are closer to the reference contours compared to the ones for the synthesized waveform using TC2. Note that since we do not perform fine-grained prosody transfer, the contours do not exactly match the ones for the reference waveform.
Figure~\ref{fig:scatter-plot} shows 16 randomly chosen references and synthesized utterances. From visual inspection, in most cases, GS-TC2 synthesized samples are closer to the reference.

\begin{table} 
\centering
	\begin{tabular}{| m{1.7cm} | c | c |}
    \hline
  	Distance Metric & TC2 & GS-TC2 \\ \hline
  	Pitch cosine & $0.052 \pm 0.005$ & $\mathbf{0.029 \pm 0.006}$ \\ \hline
  	RMS cosine &  $0.034 \pm 0.003$ & $\mathbf{0.027 \pm 0.003}$ \\ \hline 
  	Pitch DTW  & $0.540 \pm 0.035$ &  $\mathbf{0.306 \pm 0.028}$ \\ \hline 
  	RMS DTW  & $0.019 \pm 0.0007$ &  $0.019 \pm 0.001$  \\ \hline 
	\end{tabular}
	\caption{Objective metrics for prosody transfer. Lower is better.}  \label{tab:obj}
 \end{table}

\section{Conclusion and Future Work}
\label{sec:discuss}

We have demonstrated a simple yet effective method for synthesizing highly natural, expressive speech using prosody transfer. By combining acoustic features known to correlate well with prosodic attributes 
with a neural TTS synthesizer (Tacotron2), and vocoder (WaveRNN), we were able to accurately transfer global prosody from a reference speech signal to synthesized speech. The use of a low-dimensional encoding of the reference speech prevents entanglement issues experienced by more complex and larger models. 

As future work, we can consider leveraging additional components of prosody, such as phoneme durations, in the model. 
Second, in order to further assist the user's choice of reference, it would be useful to develop unsupervised methods that can identify the ``best'' set of reference speech samples from large collections of speech using only the text. 
Furthermore, pilot experiments showed that the reference audio for our model can be selected from other speakers or be comprised of para-linguistics instead of speech, provided that the statistics of the prosody features of these references are close to the statistics of the features of the target speaker. Future work includes incorporating a preprocessing module to be able to use arbitrary reference audio for prosody transfer.
Another promising future direction is using features from a learned model or using a more complex reference encoder for prosody transfer.
In pilot experiments, we trained an LSTM on logF0 and RMS contours jointly with the TC2 model for prosody transfer and found this approach to perform poorly in comparison to our method. 
Finally, we would like to devise more standardized and detailed evaluations of prosody transfer and expressive speech synthesis. 

\vfill\pagebreak

\bibliographystyle{IEEEtran}
\bibliography{refs}

\begin{thebibliography}{10}
\providecommand{\url}[1]{#1}
\csname url@samestyle\endcsname
\providecommand{\newblock}{\relax}
\providecommand{\bibinfo}[2]{#2}
\providecommand{\BIBentrySTDinterwordspacing}{\spaceskip=0pt\relax}
\providecommand{\BIBentryALTinterwordstretchfactor}{4}
\providecommand{\BIBentryALTinterwordspacing}{\spaceskip=\fontdimen2\font plus
\BIBentryALTinterwordstretchfactor\fontdimen3\font minus
  \fontdimen4\font\relax}
\providecommand{\BIBforeignlanguage}[2]{{%
\expandafter\ifx\csname l@#1\endcsname\relax
\typeout{** WARNING: IEEEtran.bst: No hyphenation pattern has been}%
\typeout{** loaded for the language `#1'. Using the pattern for}%
\typeout{** the default language instead.}%
\else
\language=\csname l@#1\endcsname
\fi
#2}}
\providecommand{\BIBdecl}{\relax}
\BIBdecl

\bibitem{skerry2018towards}
\BIBentryALTinterwordspacing
R.~Skerry-Ryan, E.~Battenberg, Y.~Xiao, Y.~Wang, D.~Stanton, J.~Shor, R.~Weiss,
  R.~Clark, and R.~A. Saurous, ``Towards end-to-end prosody transfer for
  expressive speech synthesis with tacotron,'' in \emph{Proc. 35th Int.Conf.
  Mach. Learn.}, ser. Proc. Mach. Learn. Res., vol.~80.\hskip 1em plus 0.5em
  minus 0.4em\relax PMLR, 10--15 Jul 2018, pp. 4693--4702. [Online]. Available:
  \url{http://proceedings.mlr.press/v80/skerry-ryan18a.html}
\BIBentrySTDinterwordspacing

\bibitem{theune2006generating}
M.~{Theune}, K.~{Meijs}, D.~{Heylen}, and R.~{Ordelman}, ``Generating
  expressive speech for storytelling applications,'' \emph{IEEE Trans. Audio,
  Speech, and Language Process.}, vol.~14, no.~4, pp. 1137--1144, July 2006.

\bibitem{gibiansky2017deep}
\BIBentryALTinterwordspacing
A.~Gibiansky, S.~Arik, G.~Diamos, J.~Miller, K.~Peng, W.~Ping, J.~Raiman, and
  Y.~Zhou, ``Deep voice 2: Multi-speaker neural text-to-speech,'' in
  \emph{Advances in Neural Inf. Process. Systems 30}, I.~Guyon, U.~V. Luxburg,
  S.~Bengio, H.~Wallach, R.~Fergus, S.~Vishwanathan, and R.~Garnett, Eds.\hskip
  1em plus 0.5em minus 0.4em\relax Curran Associates, Inc., 2017, pp.
  2962--2970. [Online]. Available:
  \url{http://papers.nips.cc/paper/6889-deep-voice-2-multi-speaker-neural-text-to-speech.pdf}
\BIBentrySTDinterwordspacing

\bibitem{shen2018natural}
J.~{Shen}, R.~{Pang}, R.~J. {Weiss}, M.~{Schuster}, N.~{Jaitly}, Z.~{Yang},
  Z.~{Chen}, Y.~{Zhang}, Y.~{Wang}, R.~{Skerrv-Ryan}, R.~A. {Saurous},
  Y.~{Agiomvrgiannakis}, and Y.~{Wu}, ``Natural {TTS} synthesis by conditioning
  wavenet on {MEL} spectrogram predictions,'' in \emph{IEEE Int. Conf.
  Acoustics, Speech and Signal Process.}, April 2018, pp. 4779--4783.

\bibitem{wu2018rapid}
\BIBentryALTinterwordspacing
X.~Wu, Y.~Cao, M.~Wang, S.~Liu, S.~Kang, Z.~Wu, X.~Liu, D.~Su, D.~Yu, and
  H.~Meng, ``Rapid style adaptation using residual error embedding for
  expressive speech synthesis,'' in \emph{Proc. Interspeech}, 2018, pp.
  3072--3076. [Online]. Available:
  \url{http://dx.doi.org/10.21437/Interspeech.2018-1991}
\BIBentrySTDinterwordspacing

\bibitem{zen2007hmm}
H.~Zen, T.~Nose, J.~Yamagishi, S.~Sako, T.~Masuko, A.~W. Black, and K.~Tokuda,
  ``The {HMM}-based speech synthesis system ({HTS}) version 2.0.'' in
  \emph{SSW}.\hskip 1em plus 0.5em minus 0.4em\relax Citeseer, 2007, pp.
  294--299.

\bibitem{schroder2010synthesis}
Schröder, M, Burkhardt, F, Krstulovic, and S, ``Synthesis of emotional
  speech,'' in \emph{A {Blueprint} for {Affective} {Computing}: {A}
  {Sourcebook} and {Manual}}, K.~R. Scherer, T.~Bänziger, and E.~Roesch,
  Eds.\hskip 1em plus 0.5em minus 0.4em\relax Oxford University Press, 2010.

\bibitem{eyben2012unsupervised}
F.~{Eyben}, S.~{Buchholz}, N.~{Braunschweiler}, J.~{Latorre}, V.~{Wan},
  M.~J.~F. {Gales}, and K.~{Knill}, ``Unsupervised clustering of emotion and
  voice styles for expressive {TTS},'' in \emph{IEEE Int. Conf. Acoustics,
  Speech and Signal Process.}, March 2012, pp. 4009--4012.

\bibitem{wang2017tacotron}
Y.~Wang, R.~Skerry-Ryan, D.~Stanton, Y.~Wu, R.~J. Weiss, N.~Jaitly, Z.~Yang,
  Y.~Xiao, Z.~Chen, S.~Bengio \emph{et~al.}, ``Tacotron: Towards end-to-end
  speech synthesis,'' \emph{arXiv preprint arXiv:1703.10135}, 2017.

\bibitem{wang2018style}
\BIBentryALTinterwordspacing
Y.~Wang, D.~Stanton, Y.~Zhang, R.~J. Skerry{-}Ryan, E.~Battenberg, J.~Shor,
  Y.~Xiao, Y.~Jia, F.~Ren, and R.~A. Saurous, ``Style tokens: Unsupervised
  style modeling, control and transfer in end-to-end speech synthesis,'' in
  \emph{Proc. 35th Int. Conf. Mach. Learn.}, 2018, pp. 5167--5176. [Online].
  Available: \url{http://proceedings.mlr.press/v80/wang18h.html}
\BIBentrySTDinterwordspacing

\bibitem{akuzawa2018expressive}
\BIBentryALTinterwordspacing
K.~Akuzawa, Y.~Iwasawa, and Y.~Matsuo, ``Expressive speech synthesis via
  modeling expressions with variational autoencoder,'' in \emph{Proc.
  Interspeech 2018}, 2018, pp. 3067--3071. [Online]. Available:
  \url{http://dx.doi.org/10.21437/Interspeech.2018-1113}
\BIBentrySTDinterwordspacing

\bibitem{taigman2018voiceloop}
\BIBentryALTinterwordspacing
Y.~Taigman, L.~Wolf, A.~Polyak, and E.~Nachmani, ``Voiceloop: Voice fitting and
  synthesis via a phonological loop,'' in \emph{Int. Conf. Learn.
  Representations, Conf. Track Proc.}, 2018. [Online]. Available:
  \url{https://openreview.net/forum?id=SkFAWax0-}
\BIBentrySTDinterwordspacing

\bibitem{hsu2018hierarchical}
\BIBentryALTinterwordspacing
W.-N. Hsu, Y.~Zhang, R.~Weiss, H.~Zen, Y.~Wu, Y.~Cao, and Y.~Wang,
  ``Hierarchical generative modeling for controllable speech synthesis,'' in
  \emph{Int. Conf. Learn. Representations}, 2019. [Online]. Available:
  \url{https://openreview.net/forum?id=rygkk305YQ}
\BIBentrySTDinterwordspacing

\bibitem{wan2019chive}
V.~Wan, C.-a. Chan, T.~Kenter, J.~Vit, and R.~A. Clark, ``{CHiVE}: Varying
  prosody in speech synthesis with a linguistically driven dynamic hierarchical
  conditional variational network,'' in \emph{Proc. 36th Int. Conf. Mach.
  Learn.}, 2019.

\bibitem{lee2019robust}
\BIBentryALTinterwordspacing
Y.~Lee and T.~Kim, ``Robust and fine-grained prosody control of end-to-end
  speech synthesis,'' in \emph{{IEEE} Int. Conf. Acoustics, Speech and Signal
  Process.}, 2019, pp. 5911--5915. [Online]. Available:
  \url{https://doi.org/10.1109/ICASSP.2019.8683501}
\BIBentrySTDinterwordspacing

\bibitem{Klimkov2019}
\BIBentryALTinterwordspacing
V.~Klimkov, S.~Ronanki, J.~Rohnke, and T.~Drugman, ``{Fine-Grained Robust
  Prosody Transfer for Single-Speaker Neural Text-To-Speech},'' in \emph{Proc.
  Interspeech 2019}, 2019, pp. 4440--4444. [Online]. Available:
  \url{http://dx.doi.org/10.21437/Interspeech.2019-2571}
\BIBentrySTDinterwordspacing

\bibitem{valle2019mellotron}
R.~Valle, J.~Li, R.~Prenger, and B.~Catanzaro, ``Mellotron: Multispeaker
  expressive voice synthesis by conditioning on rhythm, pitch and global style
  tokens,'' 2019.

\bibitem{Povey_ASRU2011}
D.~Povey, A.~Ghoshal, G.~Boulianne, L.~Burget, O.~Glembek, N.~Goel,
  M.~Hannemann, P.~Motlicek, Y.~Qian, P.~Schwarz, J.~Silovsky, G.~Stemmer, and
  K.~Vesely, ``The kaldi speech recognition toolkit,'' in \emph{IEEE 2011
  Workshop on Automatic Speech Recognition and Understanding}.\hskip 1em plus
  0.5em minus 0.4em\relax IEEE Signal Processing Society, Dec. 2011.

\bibitem{ghahremani2014pitch}
P.~{Ghahremani}, B.~{BabaAli}, D.~{Povey}, K.~{Riedhammer}, J.~{Trmal}, and
  S.~{Khudanpur}, ``A pitch extraction algorithm tuned for automatic speech
  recognition,'' in \emph{IEEE Int. Conf. Acoustics, Speech and Signal
  Process.}, May 2014, pp. 2494--2498.

\bibitem{kalchbrenner2018efficient}
\BIBentryALTinterwordspacing
N.~Kalchbrenner, E.~Elsen, K.~Simonyan, S.~Noury, N.~Casagrande, E.~Lockhart,
  F.~Stimberg, A.~van~den Oord, S.~Dieleman, and K.~Kavukcuoglu, ``Efficient
  neural audio synthesis,'' in \emph{Proc. 35th Int. Conf. Mach. Learn.}, 2018,
  pp. 2415--2424. [Online]. Available:
  \url{http://proceedings.mlr.press/v80/kalchbrenner18a.html}
\BIBentrySTDinterwordspacing

\bibitem{muller2007dynamic}
M.~M{\"u}ller, ``Dynamic time warping,'' \emph{Information retrieval for music
  and motion}, pp. 69--84, 2007.

\end{thebibliography}

\end{document}